\documentclass[fleqn,usenatbib]{mnras}
\usepackage[T1]{fontenc}
\usepackage{commath}
\usepackage{amsmath}	
\usepackage{amssymb}	
\usepackage{ae,aecompl}
\usepackage{graphicx}	
\usepackage[usenames,dvipsnames,svgnames,table]{xcolor}
\usepackage[table]{xcolor}
\usepackage{verbatim}
\usepackage[font = small, labelfont=bf, labelsep=period]{caption}
\usepackage{subcaption}
\usepackage{breqn}
\usepackage{float}
\usepackage{dblfloatfix}

\usepackage[utf8]{inputenc}
\usepackage{multicol}
\usepackage{bibspacing}
\usepackage{lipsum}
\usepackage{hyperref}
\usepackage{diagbox}
\usepackage[inline]{enumitem}
\usepackage{enumitem}
\usepackage[normalem]{ulem}
\setlist{
	listparindent=\parindent,
	parsep=0pt,
}
\hypersetup{ %
			colorlinks=true,%
			pdfauthor={Nomen Tuum},%
			pdftitle={Orbital Evolution of Eccentric Binaries in Disks},%
			citecolor=blue%
			}

\newcommand{\ab}{a_{\rm b}}
\newcommand{\abd}{\dot{a}_{\rm b}}
\def\red#1 {\textcolor{red}{#1}\ }  
\def \green#1 {\textcolor{green}{#1}\ }  

\title[Orbital Evolution of Binaries in Circumbinary Disks]{Orbital Evolution of Binaries in Circumbinary Disks}
\author[Magdalena Siwek et al.]{
Magdalena Siwek,$^{1}$\thanks{E-mail: magdalena.siwek@cfa.harvard.edu}
Rainer Weinberger, $^{2}$
Lars Hernquist$^{1}$
\\
$^{1}$Center for Astrophysics, Harvard University, Cambridge, MA 02138, USA \\
$^{2}$ Canadian Institute for Theoretical Astrophysics, 60 St. George St., Toronto, ON M5S 3H8, Canada \\
}

\date{Accepted XXX. Received YYY; in original form ZZZ}

\pubyear{2022}

\begin{document}
\label{firstpage}
\pagerange{\pageref{firstpage}--\pageref{lastpage}}
\maketitle

\begin{abstract}
	 We present the to-date largest parameter space exploration of binaries in circumbinary disks (CBDs), deriving orbital evolution prescriptions for eccentric, unequal mass binaries from our suite of hydrodynamic simulations. In all cases, binary eccentricities evolve towards steady state values that increase with mass ratio, and saturate at an equilibrium eccentricity $e_{\rm b, eq} \sim 0.5$ in the large mass ratio regime, in line with resonant theory. For binaries accreting at their combined Eddington limit, a steady state eccentricity can be achieved within a few Megayears. Once at their steady state eccentricities, binaries with $q_{\rm b} \gtrsim 0.3$ evolve towards coalescence, while lower mass ratio systems expand due to CBD torques. We discuss implications for population studies of massive black hole binaries, protostars in binary systems, and post-common envelope binaries observed by ground-based gravitational wave detectors.
\end{abstract}


\section{Introduction}
Circumbinary disks (CBDs) form and evolve in tandem with binaries in a range of astrophysical contexts: formation of stellar binaries in protoplanetary disks (e.g. \citealt{Dutrey1994, Mathieu1997, Tofflemire2017b, Czekala2021}), massive black hole binaries (MBHBs) immersed in AGN disks (e.g., \citealt{Begelman1980, YuTremaine2002, Vasiliev2015, Gualandris2016}), and remnants of common envelopes around progenitors of compact-object mergers (e.g., \citealt{KashiSoker2011, Reichardt2019}; see also the review by \citealt{Roepke2022}, and for a comprehensive review on circumbinary accretion disks and their applications, see \citealt{LaiMunoz2022}.)

A disk and binary interact gravitationally, resulting in: \begin{enumerate*} \item eccentricity growth in the CBD and, in some cases, precession around the binary (however, see e.g. \cite{Miranda2016} and \cite{Siwek2023a} for `locked' disks), and \item evolution of binary parameters, such as mass ratio $q_{\rm b}$ (e.g., \citealt{Artymowicz1983, Bate2002,  Farris2014, Gerosa2015, Siwek2023a}), semi-major axis $a_{\rm b}$ and eccentricity $e_{\rm b}$. The evolution of binary orbital parameters occurs due to the accretion of gas and momentum onto the binary, and the gravitational torques between binary and gas. 
\end{enumerate*}

The study of long-term orbital evolution in binaries with CBDs is a critical component to understand the properties of binary populations in current and upcoming electromagnetic (EM) and gravitational wave (GW) observatories. In this work we focus on the rate of change of the semi-major axis ($\dot{a}_{\rm b}$), and the binary eccentricity ($\dot{e}_{\rm b}$) due to gas torques. Analytical and numerical calculations of gas torques acting on binaries in CBDs show that disks can extract angular momentum from binaries, reducing the semi-major axis and assisting coalescence (e.g., \citealt{Pringle1991, Gould2000, Armitage2002,  ArmitageNatarajan2005, MacFadyen2008, Cuadra2009, Haiman2009}). Conversely, recent hydrodynamic simulations have found that the net torque acting on the binary can cause orbital expansion (e.g. \citealt{Miranda2016, Moody2019, Munoz2019, Duffell2020, Munoz2020, DorazioDuffell2021}), calling into question whether CBDs contribute to coalescence, or prevent binaries from merging. However, a large parameter study including eccentric, unequal mass binaries has not yet been done, and is the focus of this work.


Dynamical interaction with gas also changes the orbital eccentricity of the binary. Analytical calculations suggest that Lindblad resonances cause eccentricity excitation, while co-rotation resonances circularize the binary orbit \citep{Goldreich1980, Artymowicz1991}.  \cite{LubowArtymowicz1992} found that the balance between Lindblad and co-rotation resonances evolves binaries with small initial eccentricity and mass ratios above $q_{\rm b} \gtrsim 0.2$ to orbital eccentricities close to $e_{\rm b} \sim 0.5$. Similarly, more recent hydrodynamic simulations have found that binaries with equal mass ratios tend towards an `equilibrium eccentricity' $e_{\rm b, eq} \sim 0.45$ \citep{Zrake2021, DorazioDuffell2021}. In our work we include the lower mass ratio regime, establishing whether an equilibrium eccentricity $e_{\rm b, eq}$ exists for all binaries in the range $0.1 \leq q_{\rm b} \leq 1.0$, and the relationship between $e_{\rm b, eq}$ and $q_{\rm b}$.


In \cite{Siwek2023a} we provided a detailed analysis of the preferential accretion rates onto binaries of various mass ratios and eccentricities. In this follow-up work, we further expand the parameter space, including more values of $e_{\rm b}$, and calculate the change in binary semi-major axis and eccentricity as a function of the binary parameters.
We study the torques exerted by the gas in the CBD and the circumsingle disks (CSDs) on the binary, both through accretion and gravitational interaction, for a large parameter space covering binary mass ratio and eccentricity. We further include an analysis that explores the contribution of torques from the CBD region only, excluding the cavity. 

This paper is structured as follows. In Section \ref{sec:numerical} we present the initial conditions and numerical methods used to evolve our hydrodynamic simulations, including the binary orbital evolution calculations (Section \ref{sec:torques}). In Section \ref{sec:results} we present results from our simulations. Our main result is a study of gravitational and accretion torques as a function of $e_{\rm b}$ and $q_{\rm b}$, and resulting evolution of binary semi-major axis and eccentricity.
In Section \ref{sec:discussion} we discuss our results in a broader astrophysical context. We outline implications for both EM and GW observations of binary populations, including binary stars and compact objects. 

\section{Numerical Methods}
\label{sec:numerical}
We carry out hydrodynamic simulations of binaries immersed in CBDs using the moving mesh code Arepo \citep{Springel2010, Pakmor2016} in its Navier Stokes version \citep{Munoz2013}, which employs a Voronoi tessellation to generate a moving grid around a set of discrete mesh-generating points. 

Similar to the simulation suite presented in \cite{Siwek2023a}, we initialize our simulations with a binary represented by two sink particles with a fixed mass ratio $q_{\rm b} \equiv \frac{M_2}{M_1}$, where $M_2$ and $M_1$ are the secondary and primary binary component respectively. We scale the simulation by choosing the total binary mass $M_{\rm b} = M_1 + M_2 \equiv 1$. Each binary component is represented by a sink particle with sink radius $r_{\rm s} = 0.03 a_{\rm b}$, where $a_{\rm b}$ is the binary semi-major axis. 
Within the sink region, we remove a fraction of gas from the gas cells at each time step. The fraction of gas removed from gas cells depends on the distance from the sink, and is defined by the dimensionless parameter $\gamma = \gamma_0 \times (1 - r_{\rm ij}/r_{\rm s})^2$, where $\gamma_0 = 0.5$ and $r_{\rm ij}$ is the distance between the j-th sink particle and i-th gas cell. 

The sink particles move on a fixed Keplerian orbit with semimajor axis $a_{\rm b}$ and eccentricity $e_{\rm b}$, which are both fixed in time. When measuring the binary orbital evolution in this work, we calculate the specific angular momentum and energy rates of change of the binary, and thus the inferred change in the orbital parameters (see equations \ref{eqn:dotab} and \ref{eqn:doteb} in section \ref{sec:torques}). We do not measure the orbital evolution directly from the simulation by using `live' orbits. We choose this method \begin{enumerate*} \item  to avoid numerical errors due to orbit integration, and \item to allow the binary and disk to achieve a steady state before measuring the orbital evolution.\end{enumerate*} 

The simulation setup is identical to \cite{Siwek2023a}. The circumbinary accretion disk is modeled as a finite, locally isothermal torus with an $\alpha$-viscosity, where $\alpha = 0.1$ is a constant. Throughout, we choose an aspect ratio $h = 0.1$. 
The binary-disk system is placed in a 2D computational box of size $300a_{\rm b} \times 300 a_{\rm b}$ with open boundary conditions, allowing the disk to viscously spread over an integration time of $10\, 000$ binary orbital timescales ($P_{\rm b}$). 

Since our numerical methods are identical to those previously presented, more details can be found in  \cite{Siwek2023a}, with the addition of the orbital evolution calculations outlined in this work's section \ref{sec:torques}.

\subsection{Sink Radius Study}
\label{sec:convergence_methods}
In Section \ref{sec:convergence_high_el}  we investigate the influence of sink sizes on the measured orbital evolution. We test sink radii in the range $r_{\rm s} = [0.03\,a_{\rm b}, 0.01\,a_{\rm b}, 0.005\,a_{\rm b}]$, with the largest as our fiducial value. Since simulations with very small sink radii are prohibitively expensive, we do not evolve each simulation for $10\,000\, P_{\rm b}$, as we do in our fiducial simulation suite. Instead, we take a snapshot from our fiducial simulation at $t=3000\,P_{\rm b}$ for binaries with equal mass ratios and eccentricities $e_{\rm b} = [0.0, 0.2, 0.4, 0.6, 0.8]$ and restart the simulation with a smaller sink radius. We do this to allow the CBD to viscously relax first, which reduces the required run-time of the simulation with smaller, and more computationally expensive, sink radii. We then evolve this simulation for an additional $2000\,P_{\rm b}$, over which we measure the orbital evolution based on the new sink parameters.

\subsection{Numerical integration of gas torques}
\label{sec:torques}
We directly compute the specific angular momentum $\delta \textbf{l}_{\rm b}$ and specific energy $\delta \epsilon_{\rm b}$ deposited into our sinks in each timestep $\delta t$:

\begin{equation}
	\delta \textbf{l}_{\rm b} = (\textbf{r}_{\rm b}  \times \textbf{f}_{\rm ext} ) \,\delta t,
\label{eqn:lbdot}
\end{equation}
\begin{equation}
	\delta \epsilon_{\rm b} = (\textbf{v}_{\rm b} \cdot  \textbf{f}_{\rm ext})  \,\delta t,
\label{eqn:epsbdot}
\end{equation}
where $\textbf{r}_{\rm b} = \textbf{r}_{1} - \textbf{r}_{2}$ and $\textbf{v}_{\rm b} = \textbf{v}_{1} - \textbf{v}_{2}$ are the relative positions and velocities of the binary, respectively. 
The external forces $\textbf{f}_{\rm ext}$ acting on the sinks include gravitational (section \ref{sec:gravitational_torques}) and accretion forces (section \ref{sec:accretion_torques}),
\begin{equation}
\textbf{f}_{\rm ext} = (\textbf{f}_{\rm g, 1} - \textbf{f}_{\rm g, 2})  + (\textbf{f}_{\rm acc, 1} - \textbf{f}_{\rm acc, 2}),
\label{eqn:f_ext}
\end{equation} 
where $\textbf{f}_{\rm g, i}$ and $\textbf{f}_{\rm acc, i}$ are defined in equations \ref{eqn:fgrav} and \ref{eqn:facc} respectively.
Binary eccentricity $e_{\rm b}$ and semi-major axis $a_{\rm b}$ evolve due to the change in specific energy and angular momentum. We compute the rate of change of eccentricity $\dot{e}_{\rm b}$ and  semi-major axis $\dot{a}_{\rm b}$ (see e.g. \citealt{Munoz2019,DorazioDuffell2021}),

\begin{equation}
\dot{e}_{\rm b} =  -\frac{1-e_{\rm b}^2}{e_{\rm b}^2}\Big[2\,\frac{\dot{M}_{\rm b}}{M_{\rm b}} - \frac{\dot{\epsilon}_{\rm b}}{\epsilon_{\rm b}} - 2\, \frac{\dot{l}_{\rm b}}{l_{\rm b}}\Big], 
\label{eqn:doteb}
\end{equation}

\begin{equation}
\frac{\abd}{\ab} = \frac{\dot{M}_{\rm b}}{M_{\rm b}} - \frac{\dot{\epsilon}_{\rm b}}{\epsilon_{\rm b}}.
\label{eqn:dotab}
\end{equation}
The continuous rate of change of specific angular momentum $\dot{l}_{\rm b}$ and energy $\dot{\epsilon}_{\rm b}$ is calculated by taking the finite difference of equations \ref{eqn:lbdot} and \ref{eqn:epsbdot}.
We time average the changes in specific energy and angular momentum to calculate the binary orbital evolution as in equations \ref{eqn:doteb} and \ref{eqn:dotab} over the simulation time of $10\,000\,P_{\rm b}$.

\subsubsection{Gravitational Torques}
\label{sec:gravitational_torques}
Gravitational interaction with the surrounding gas cells results in a gravitational acceleration $\textbf{f}_{\rm g, i}$ of the i-th sink particle,
\begin{equation}
\textbf{f}_{\rm g, i} = -\mathcal{G} \sum_{\rm j} m_j \frac{(\textbf{r}_{\rm i} - \textbf{r}_{\rm j})}{\left| \textbf{r}_{\rm i} - \textbf{r}_{\rm j}\right |^3}.
\label{eqn:fgrav}
\end{equation}
Here, sink particle $i$ is at a location denoted by $\textbf{r}_{\rm i}$, the position vector of the sink particle as measured from the simulation barycenter, $m_{\rm j}$ is the mass and $\textbf{r}_{\rm j}$ the position of the j-th gas cell, and $\mathcal{G}$ is the gravitational constant.

\subsubsection{Accretion Torques}
\label{sec:accretion_torques}
Gas cells within the sink radius $r_{\rm s}$ are drained by a factor $\gamma$ in each timestep (see \cite{Siwek2023a} for more details on our numerical methods). In addition to mass, linear momentum is accreted by the sink, resulting in an asymmetric accretion force onto the sink particle within each timestep \citep{Roedig2012, Munoz2019}. The accretion force during a timestep $\delta t$ due to accretion of mass and momentum from a cell with index i takes the form,

\begin{equation}
	\textbf{f}_{\rm acc, i} = \Big(\frac{\delta \textbf{p}_{\rm i} - \delta m_{\rm i} \textbf{v}_{\rm i}}{m_{\rm i} + \delta m_{\rm i}}\Big)/\delta t.
	\label{eqn:facc}
\end{equation}

Here $\delta \textbf{p}_{\rm i}$ and $\delta m_{\rm i}$ are the momentum and mass accreted by sink particle i in the current timestep, while $m_{\rm i}$ is the sink particle mass prior to the current timestep. $\textbf{v}_{\rm i}$ is the velocity vector of the sink particle, and $\delta t$ is the current timestep of the simulation. 
Summing over all gas cells which are drained per timestep yields the specific force due to accretion acting on each sink per timestep. 

\section{Results}
\label{sec:results}

\subsection{Semi-major axis evolution}
\label{sec:semi-major_evol}
\begin{figure*}
	\centering
	\includegraphics[width=1.0\textwidth]{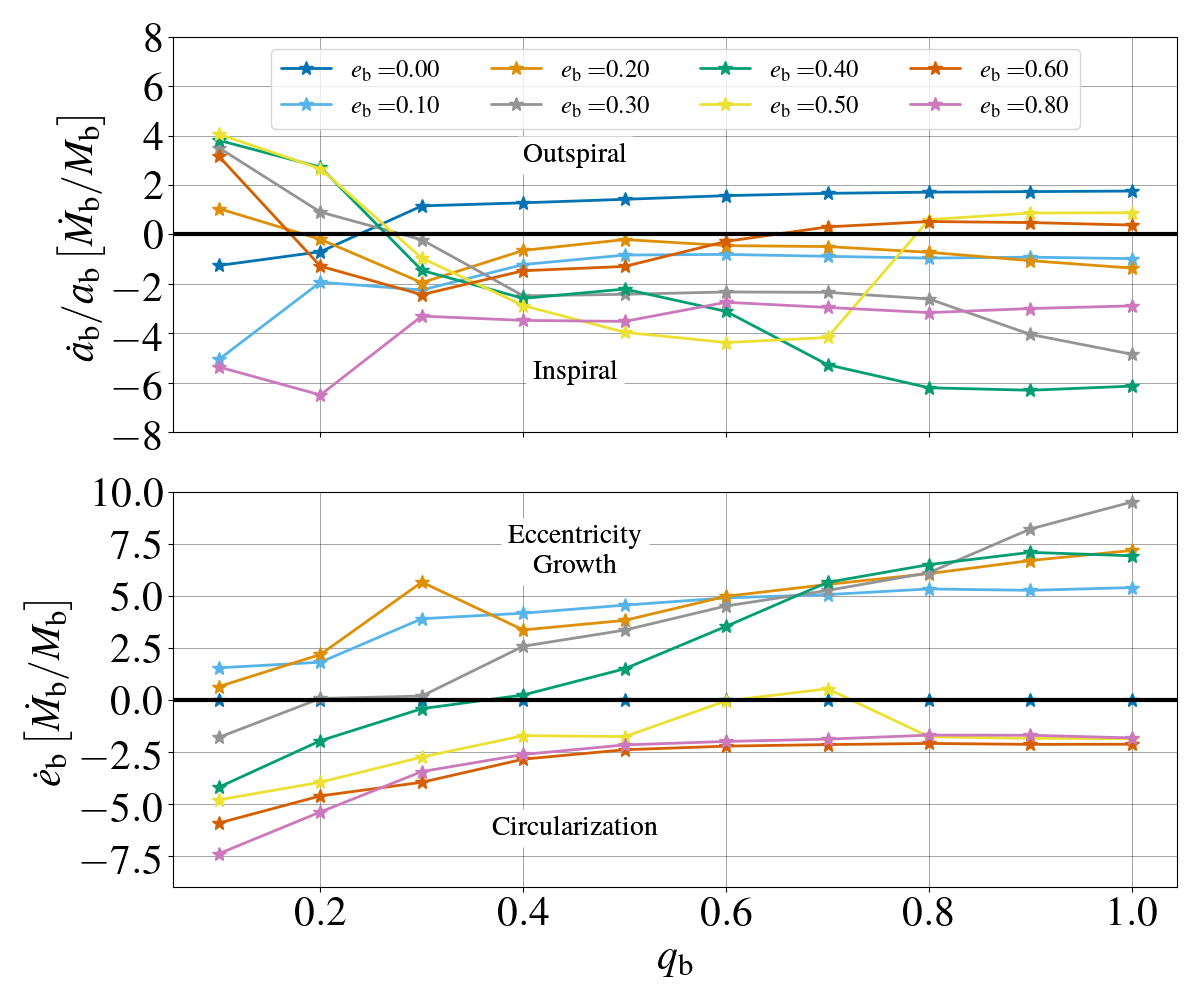}
	\caption{Semi-major axis (top) and eccentricity (bottom) evolution of binaries in our simulation suite, including gravitational and accretion forces. We find that most binaries above $q_{\rm b} > 0.2$ migrate outward,with the exception of a few cases with eccentricities $e_{\rm b} = 0.5$ and $e_{\rm b} = 0.6$. Eccentricity growth increases as a function of mass ratio, with all mass ratios experiencing regimes of circularization and eccentricity growth.}
	\label{fig:dotab_doteb_grav_acc}
\end{figure*}

We calculate the evolution of the binary semi-major axis as a function of $q_{\rm b}$ and $e_{\rm b}$ as in equation \ref{eqn:dotab}. We report the mean of $\dot{a}_{\rm b}/a_{\rm b}$ over the last 7000 orbits, after the disk is viscously relaxed out to $\gtrsim 10\, a_{\rm b}$. We show $\dot{a}_{\rm b}/a_{\rm b}$  for each one of our 80 simulations, and plot the result as a function of mass ratio in the top panel of Figure \ref{fig:dotab_doteb_grav_acc}. For circular binaries, we find that the semi-major axis expands for all binaries with $q_{\rm b} \gtrsim 0.2$ (previously shown in \citealt{Munoz2020}). However, in most binaries with non-zero eccentricites (up to $e_{\rm b} = 0.8$ in this work), we find that CBD torques lead to inspiral. We note some exceptions in low mass ratio, eccentric binaries ($q_{\rm b} \lesssim 0.2, \, 0.2 \lesssim e_{\rm b} \lesssim 0.6 $) and several high mass ratio binaries with $e_{\rm b} = 0.5$ and $e_{\rm b} = 0.6$, where $\dot{a}_{\rm b}$ is just greater than 0. 

\begin{figure*}
	\centering
	\includegraphics[width=1.0\textwidth]{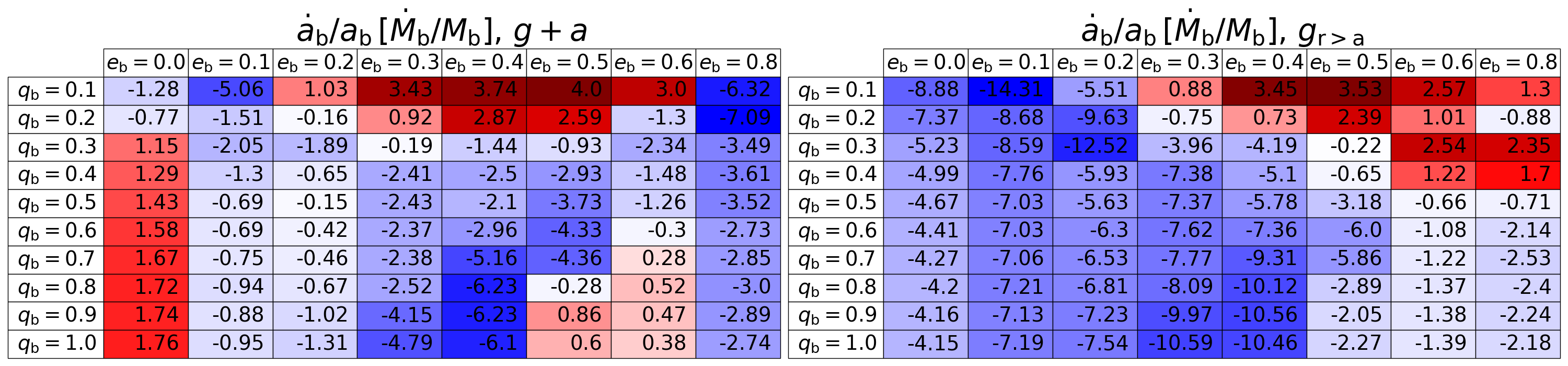}
	\caption{Tables showing the semi-major axis rate of change based on gravitational and accretion torques (left) or gravitational torques from the outer CBD region (right; at a distance $r_{\rm cell} > a_{\rm b}$ from barycenter), for all binaries in our parameter study. 
	The colourmap is centered around 0, with negative $\dot{a}_{\rm b}$values in blue (`inspiral'), and positive $\dot{a}_{\rm b}$ values in red (`outspiral'). }
	\label{fig:table_abdot}
\end{figure*}

The fiducial inspiral/outspiral regimes are shown again in Figure \ref{fig:table_abdot} (left table), where we present numerical values of the orbital migration rate $\dot{a}_{\rm b}$ for each simulation. Blue indicates negative values of  $\dot{a}_{\rm b}$, implying that the binary shrinks, while red indicates positive values of $\dot{a}_{\rm b}$, implying that the binary expands. The majority (58 out of our 80 simulations) of binaries we tested exhibit negative migration rates, suggesting that CBD torques assist binary coalescence in the majority of the parameter space sampled here. Significant positive migration rates are mostly seen when:
\begin{enumerate*}
	\item the binary moves on a circular orbit, and
	\item in systems with small mass ratios. 
\end{enumerate*}
We note that the semi-major axis expansion in circular binaries originates in the cavity/CSD region. In the right panel of Figure \ref{fig:table_abdot}, we show the semi-major axis evolution of binaries due to gas torques from cells at a distance $r \geq a_{\rm b}$ from the binary barycenter. We calculate these gas torques by excising the cavity region $r < a_{\rm b}$ at every timestep, when calculating the gravitational forces acting between the binary and gas cells. When comparing the torques acting on circular binaries, in both tables, we find that torques from gas cells in the outer CBD lead to binary inspiral, as indicated by the blue colours in the left-most column in Figure \ref{fig:table_abdot} (right hand side table). In the absence of the contribution from the inner region (including the CSDs), torques from the CBD would give rise to binary inspiral, as has been traditionally argued in the literature (e.g., \citealt{Pringle1991, Gould2000, Armitage2002}). However, in circular binaries, the positive contribution from the CSDs outweighs this effect, leading to the now well documented phenomenon of binary outspiral.

The origin of the orbital expansion and hardening rates in eccentric binaries is less clear. In eccentric low mass ratio binaries, significant positive torques remain when removing the cavity region from the torque calculation, as seen in the right hand side of Figure \ref{fig:table_abdot}. This could indicate that torques from the CBD may contribute to binary expansion, if the binary is eccentric. However, we note that the $r \leq a_{\rm b}$ region we remove from the torque calculation may be too small to entirely exclude the cavity region in eccentric, low mass ratio binaries. This is because the cavity size increases with increasing eccentricity due to the changing locations of resonances (see, e.g., \cite{Artymowicz1994} for a study on gap sizes in CBDs). Streams and CSDs around the secondary may therefore not be excised entirely, and may still contribute to the orbital evolution calculation. The expansion rates shown in low mass ratio, highly eccentric binaries may therefore originate in the CSD of the secondary as well as the CBD. 

Nevertheless, in circular binaries, and eccentric equal mass binaries, the origin of the torques causing expansion is in the CSD, consistent with predictions by \cite{GoldreichTremaine1980}, and recently shown with hydrodynamic simulations in \cite{Munoz2020}. 

\begin{figure*}
	\centering
	\includegraphics[width=1.0\textwidth]{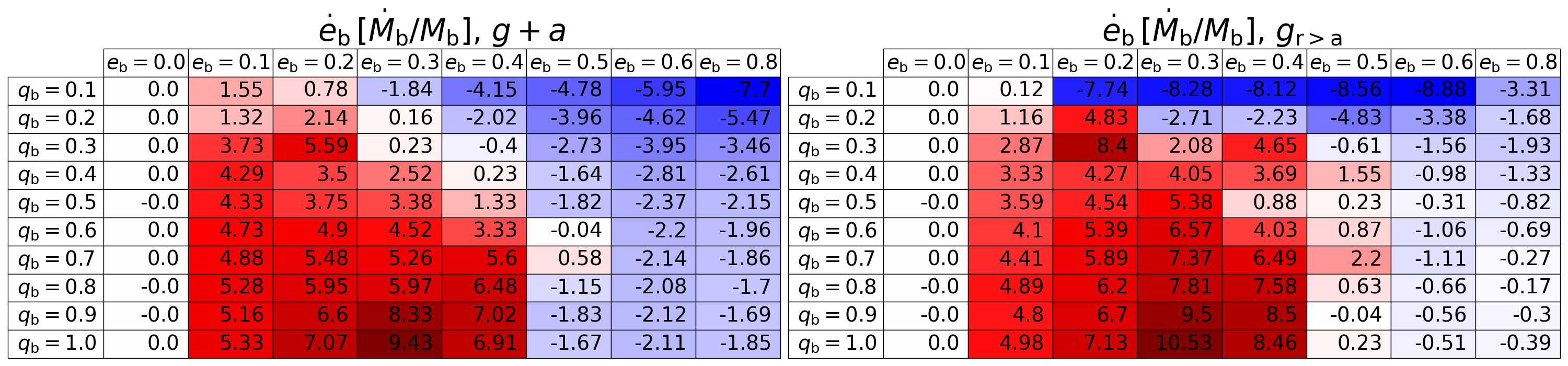}
	\caption{Table showing the binary eccentricity rate of change based on gravitational and accretion torques (left) or gravitational torques from the outer CBD region (right; at a distance $r_{\rm cell} > a_{\rm b}$ from barycenter), for all binaries in our parameter study. The colourmap is centered around 0, with negative $\dot{e}_{\rm b}$ values in blue (`circularization'), and positive $\dot{e}_{\rm b}$  values in red (`eccentricity growth').}
	\label{fig:table_ebdot}
\end{figure*}

\subsection{Eccentricity evolution}
\label{sec:eccentricity_evol}
Binary eccentricity evolves as a result of interaction with the CBD (e.g. \citealt{Roedig2011, Miranda2016, Munoz2019, Dorazio2021}), however the parameter space explored has been mostly limited to equal mass binaries. Here we show how the orbital eccentricity of eccentric, (mostly) unequal mass ratio binaries evolves in our simulations. The bottom panel of Figure  \ref{fig:dotab_doteb_grav_acc} shows the rate of change of binary eccentricity $\dot{e}_{\rm b}$ as a function of mass ratio, including contributions from both gravitational and accretion torques. Specific numerical values for each simulation are given in Figure \ref{fig:table_ebdot}. Table cells highlighted in red indicate eccentricity growth, while blue indicates eccentricity damping.

We note that the rate of change of the orbital eccentricity $\dot{e}_{\rm b}$ generally increases as a function of mass ratios, for a given binary eccentricity.
In addition, in each mass ratio bin, a transition from eccentricity growth ($\dot{e}_{\rm b} > 0$) to eccentricity damping ($\dot{e}_{\rm b} <  0$) takes place.  This transition can be seen in Figure \ref{fig:table_ebdot}, where red cells indicate eccentricity growth, and blue cells indicate eccentricity damping. An off-white diagonal through the parameter space separates red from blue cells, and shows the transition between the two regimes: this transition indicates the existence of an equilibrium eccentricity for binaries of all mass ratios. This equilibrium eccentricity can be calculated for each simulation, and is defined as the eccentricity at which $\dot{e}_{\rm b} \sim 0$, representing the steady state of the binary system. 

\begin{figure*}
	\centering
	\includegraphics[width=1.0\textwidth]{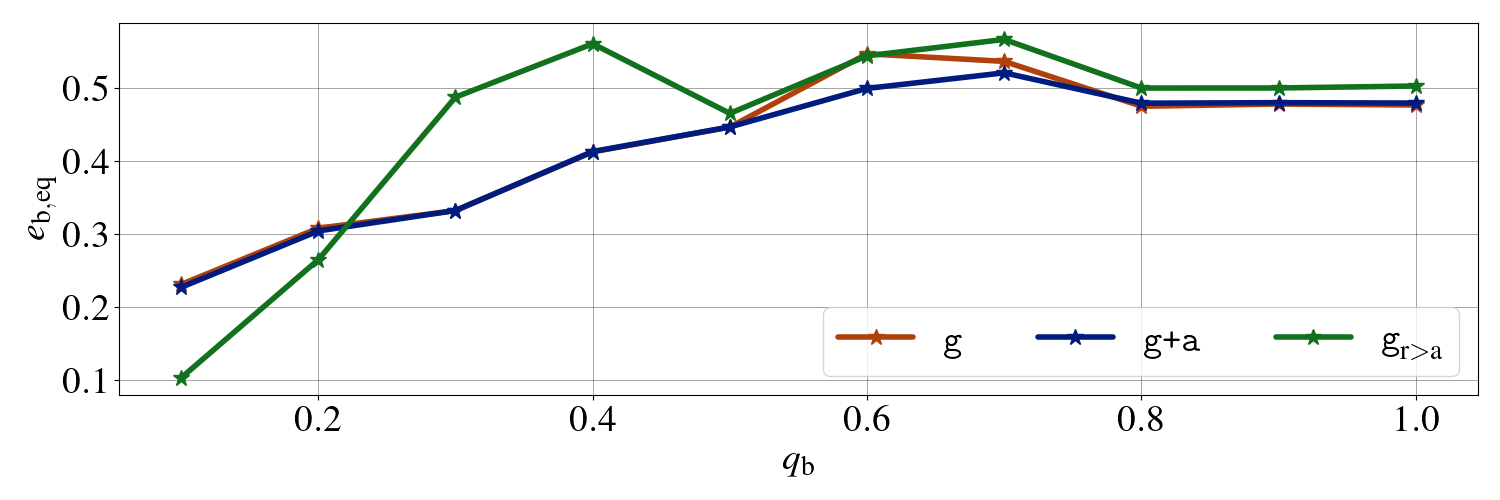}
	\caption{The `equilibrium eccentricity', $e_{\rm b, eq}$, as a function of binary mass ratio, for our 3 methods of torque calculations.  $e_{\rm b, eq}$ grows approximately linearly as a function of mass ratio, saturating when $q_{\rm b} \gtrsim 0.6$.}
	\label{fig:doteb_vs_eb}
\end{figure*}

In Figure \ref{fig:doteb_vs_eb} we show the equilibrium eccentricity of each binary as a function of $q_{\rm b}$, for 3 different torque calculations: our fiducial case including gravitational and accretion torques ($\textbf{g+a}$, blue line), a test case including gravitational torques only ($\textbf{g}$, red line), and a third case including only gravitational torques from the CBD at a distance $r \geq a_{\rm b}$ from the barycenter ($\textbf{g}_{\rm r>a}$, green line). These three cases are shown to:
\begin{enumerate*}
	\item investigate the impact of accretion torques on orbital eccentricity evolution, and
	\item distinguish between torques acting on the binary from the CBD (outer region) and cavity (including CSDs and gas streams).
\end{enumerate*}

In our fiducial case (blue line), we find that low mass ratio binaries in our simulations tend towards low-moderate equilibrium eccentricities $e_{\rm b,eq} \sim 0.2$. As the binary mass ratio grows, the equilibrium eccentricity increases at an approximately linear rate, until the growth of $e_{\rm b,eq}$ with mass ratio saturates at $q_{\rm b} \sim 0.6$, settling at around $e_{\rm b,eq} \sim 0.5$. In binaries with equal mass ratio, the orbital eccentricity evolves towards a steady state value $e_{\rm b,eq} \sim 0.48$, similar to recent results in the literature (compare with \citealt{Munoz2019, Zrake2021, DorazioDuffell2021}). 
Our main result is that $e_{\rm b,eq}$ grows with $q_{\rm b}$, so that binaries with higher mass ratios evolve towards higher equilibrium eccentricities.

We find this behaviour to be the same whether accretion torques are included in the calculation or not. The red line in Figure \ref{fig:doteb_vs_eb} represents the orbital eccentricity towards which binaries evolve when including only gravitational torques in our calculations.  Aside from a small deviation near $q_{\rm b} \sim 0.6$, the $\textbf{g+a}$ (blue line)  and $\textbf{g}$ (red line) calculations yield nearly identical equilibrium eccentricities as a function of mass ratio. This implies that accretion torques, which act only in the near vicinity of the sink particles, play a subdominant role in the eccentricity evolution of binaries immersed in CBDs. Instead, the gravitational interaction with gas in the CBD, rather than the cavity region (including the CSDs), is likely more important.

We investigate the origin of the eccentricity rate of change further by calculating $\dot{e}_{\rm b}$ including only torques from gas in the CBD region, i.e. excising any cells within a radius $r \leq a_{\rm b}$. By doing so, we exclude contributions from the cavity and CSDs. We find that the $\textbf{g}_{\rm r>a}$  calculation (green line in Figure \ref{fig:doteb_vs_eb}) yields a similar result to the other two cases: at low mass ratio, $e_{\rm b,eq}$ grows before flattening at a mass ratio $q_{\rm b} \gtrsim 0.3$, towards an equilibrium eccentricity $e_{\rm b,eq} \gtrsim 0.5$. Our results indicate that the equilibrium eccentricity of the binary is predominantly set by interactions with gas in the CBD, as opposed to the CSDs. While the CSDs play a crucial role in the evolution of the semi-major axis, they seem to be less important in the evolution of the binary eccentricity. This result suggests, in line with previous work by \cite{Artymowicz1991, LubowArtymowicz1992}, that the eccentricity growth and damping is caused by resonant interactions between the binary and the CBD. We expand the discussion on this topic in section \ref{sec:discussion}. 

\subsection{Evolution timescales}
\label{sec:timescales}
\begin{figure}
	\centering
	\includegraphics[width=1.0\columnwidth]{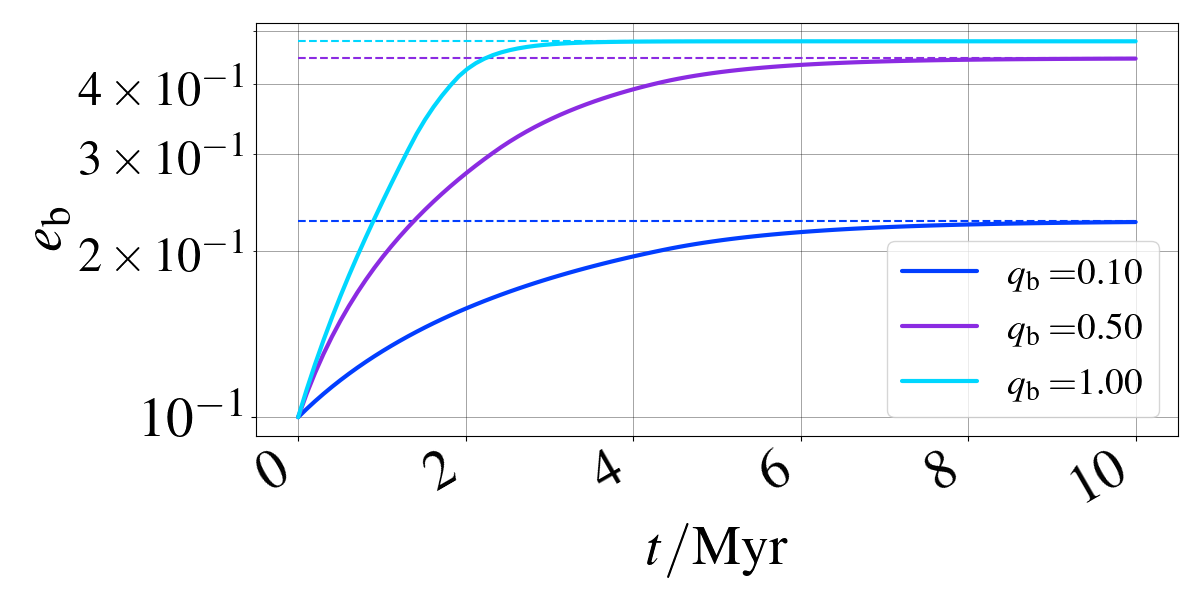}
	\caption{Eccentricity evolution as a function of time in Myrs, starting at an initial eccentricity $e_{\rm b,0} = 0.1$ and evolving towards the equilibrium eccentricity $e_{\rm b, eq}$ within $\lesssim 8 $Myrs in all cases shown. All binaries are accreting at the combined Eddington limit for both objects.}
	\label{fig:eb_vs_t}
\end{figure}
In Figure \ref{fig:eb_vs_t} we show the eccentricity evolution as a function of time, including 3 binaries with mass ratios $q_{\rm b} = 0.1, \, 0.5, \, 1.0$. The eccentricity evolution is calculated using semi-analytic evolution in post-processing: We initialize each binary with its mass ratio and initial eccentricity $e_{\rm b,0} = 0.1$, and evolve the eccentricity forward in time by interpolating $\dot{e}_{\rm b}$ from Figure \ref{fig:dotab_doteb_grav_acc}. Throughout, we assume that the binary is accreting at its Eddington limit, with a radiative efficiency $\epsilon = 0.1$. We find the timescale to reach the equilibrium eccentricity to within a few percent is $\lesssim 5 \, \rm{Myr}$, in all 3 cases shown. For reference, the Salpeter timescale at $10\%$ radiative efficiency is $t_{\rm S} \sim 45 \, \rm{Myr} $.

\subsection{Comparison with  Literature}
\begin{figure*}
	\centering
	\includegraphics[width=1.0\textwidth]{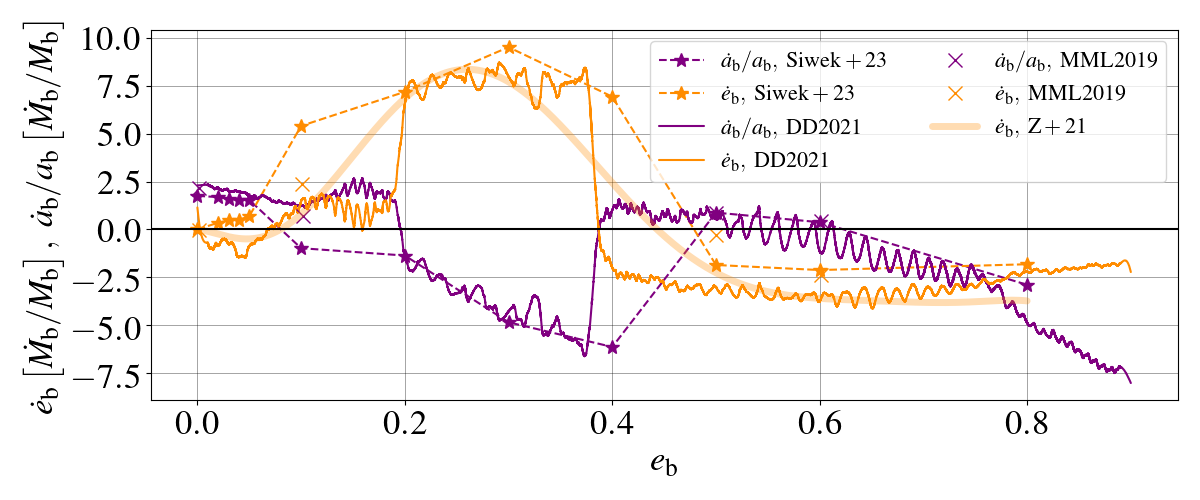}
	\caption{Literature comparison of hydrodynamic simulations of CBDs around equal mass binaries with varying eccentricities, measuring the rates of change of semi-major axis (purple markers/lines) and binary eccentricity (orange markers/lines). We include work from  \protect\cite{Munoz2019} (purple/orange crosses), a fitting function of eccentricity rate of change from \protect\cite{Zrake2021} (faint orange line), and orbital evolution results from \protect\cite{DorazioDuffell2021} (solid purple/orange lines). We find that both eccentricity and semimajor axis evolution mostly agree between different studies, with some qualitative disagreement at $ e_{\rm b} \sim 0.1$. All studies find similar equilibrium eccentricities for $q_{\rm b} = 1.0$ binaries in the range $0.4 \lesssim e_{\rm b, eq} \lesssim 0.5$.}
	\label{fig:literature_comparison}
\end{figure*}

Similar to Figure 1 in \cite{DorazioDuffell2021}, we compare the orbital evolution of equal mass ratio binaries as a function of binary eccentricity to existing results in the literature, shown here in Figure \ref{fig:literature_comparison}. We show the rate of change of the semi-major axis $\dot{a}_{\rm b}$ and binary eccentricity $\dot{e}_{\rm b}$ as a function of the fixed binary eccentricity. Our simulations are represented by orange and purple stars ($\dot{a}_{\rm b}/a_{\rm b}$ and $\dot{e}_{\rm b}$ respectively), connected by dashed lines. 
The results are in good agreement with previous simulations of equal mass, eccentric binaries (e.g. \citealt{Munoz2019, Zrake2021, DorazioDuffell2021}), with some exceptions at low eccentricity (e.g. near $e_{\rm b} \sim 0.1$). Our simulations do not show a circularization regime at $e_{\rm b} \lesssim 0.05$. Instead, equal mass ratio binaries grow more eccentric, even with only small initial eccentricities $e_{\rm b} \gtrsim 0.01$. This difference could be due to the viscosity in our simulations: we use a viscosity model that varies with radius, adopting a constant $\alpha = 0.1$, while  \cite{DorazioDuffell2021} choose a constant coefficient of kinematic
viscosity. The eccentricity growth and damping are governed by resonances in the CBD, and the strength to which each resonance is excited depends on the mass density of gas at the resonance location \citep{LubowArtymowicz1992}. Therefore, the viscosity model could play a role in the balancing of the resonances, affecting the eccentricity growth and damping slightly. This could explain the minor differences between our work and previous studies shown in Figure \ref{fig:literature_comparison}.

All simulations agree that there is an ``equilibrium eccentricity" in the range $0.4 \lesssim e_{\rm b} \lesssim 0.5$, towards which most binaries evolve. Despite some minor differences in the low eccentricity regime, the simulations agree qualitatively. We point out that the comparison in Figure \ref{fig:literature_comparison} includes results from 3 different hydrodynamic codes: both this work and \cite{Munoz2019} use \texttt{Arepo} \citep{Springel2010}, while \cite{Zrake2021} is based on \texttt{Mara3} \citep{ZrakeMacFadyen2012} and \cite{DorazioDuffell2021} on \texttt{DISCO} \citep{Duffell2016}. The qualitative agreement between different hydrodynamic codes indicates that both the value and existence of an equilibrium eccentricity in equal mass ratio binaries is robust.

\subsection{Dependence on Sink Radii \& Accretion Torques}
\label{sec:convergence_high_el}
Sink particles are commonly used to model accretion of gas in CBD simulations, however the detailed sink prescriptions, including sink rate and sink radius, may change the outcome of the simulation (e.g., \citealt{Dittmann2021b}). As part of our parameter study, we test the effect of the sink radius on the measured orbital evolution of the binary. 

Binaries with high eccentricities approach small separations at pericenter, during which the CSDs surrounding the sink particles may be tidally disrupted. Whether or not the innermost resolved region of the accretion disk around the sink particle is stripped depends on the sink radius $r_{\rm s}$ size relative to the Eggleton-Roche radius  \citep{Eggleton1983},
\begin{equation}
	r_{\rm ER} = \frac{0.49 q_{\rm b}^{2/3}}{0.6q_{\rm b}^{2/3}+\log{\Big[1+q_{\rm b}^{1/3}\Big]}} \times  r_{\rm b},
	\label{eqn:ER}
\end{equation}
where $r_{\rm b} = a_{\rm b} (1 - e_{\rm b})$ is the binary separation at pericenter. In our fiducial simulations, the sink radius $r_{\rm s} = 0.03 a_{\rm b}$ becomes comparable to $r_{\rm ER} $ when $e_{\rm b} = 0.8$, and as a result, the accretion disks around sink particles on orbits with $e_{\rm b} = 0.8$ are likely to experience distortions due to tidal forces at pericentric approach. 

As CSDs are tidally compressed or stretched during pericentric approach, the orbital motion of the gas is no longer Keplerian, and consequently, the accretion onto the sink is no longer isotropic. Accreted gas may enter the sink region from preferred directions, depending on the tidal distortion of the CSD and the motion of the sink. Anisotropic accretion torques could then become important, as the innermost region of the accretion disk around the sink particle is no longer accreting isotropically. In extreme cases, if the CSD is entirely stripped, the sink particle moves through a `headwind'  when it directly encounters gas streams in the cavity region, leading to additional anisotropic accretion. 

As the tidal distortions in the CSD move from the outer edge of the disk inwards, the size of the sink likely determines the magnitude of the resulting anisotropic accretion torques. Here we present the results of a sink radius parameter study, determining the deviations in orbital evolution due to anisotropic accretion torques, given a range of sink radii.  

\begin{figure*}
	\centering
	\includegraphics[width=1.0\textwidth]{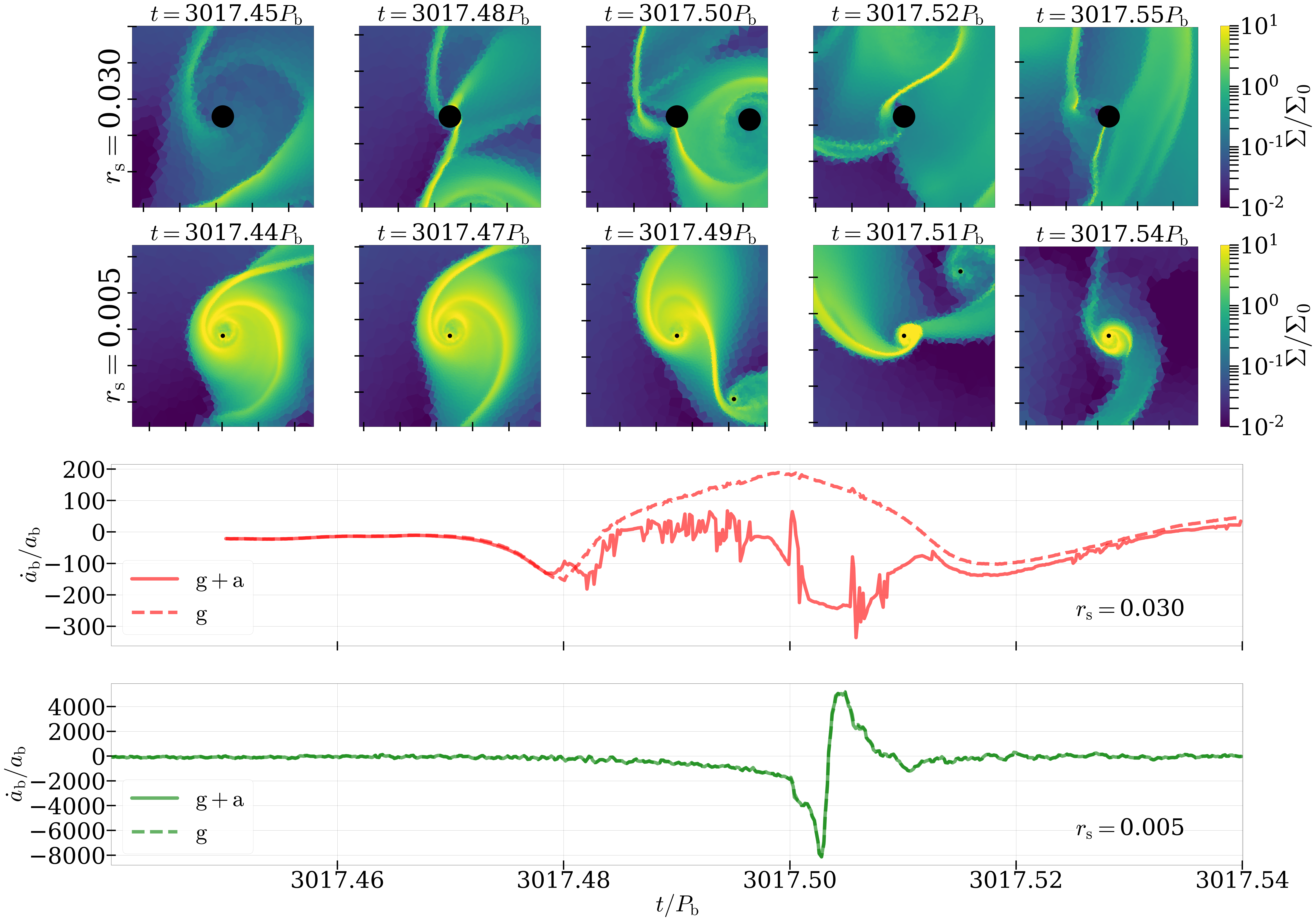}
	\caption{Top panels: successive surface density snapshots centered around the secondary in a $e_{\rm b} = 0.8, \, q_{\rm b} = 0.1$ simulation with sink radii $r_{\rm s} = 0.03 \, a_{\rm b}$ (top row) and $r_{\rm s} = 0.005 \, a_{\rm b}$ (second row). We find that the largest sink particle loses its disk structure at pericenter, while the CSD around the smallest sink particle is never entirely stripped. The bottom two panels show a time series of semi-major axis rate of change in all three simulations, comparing the orbital evolution when including gravitational torques only (\texttt{g} in the legend, dashed lines), and including anisotropic accretion torques (\texttt{g+a}, solid lines). We find that \texttt{g} and \texttt{g+a} calculations converge in simulations with smaller sink radii.}
	\label{fig:snapshots_adot}
\end{figure*}

\begin{figure*}
	\centering
	\includegraphics[width=1.0\textwidth]{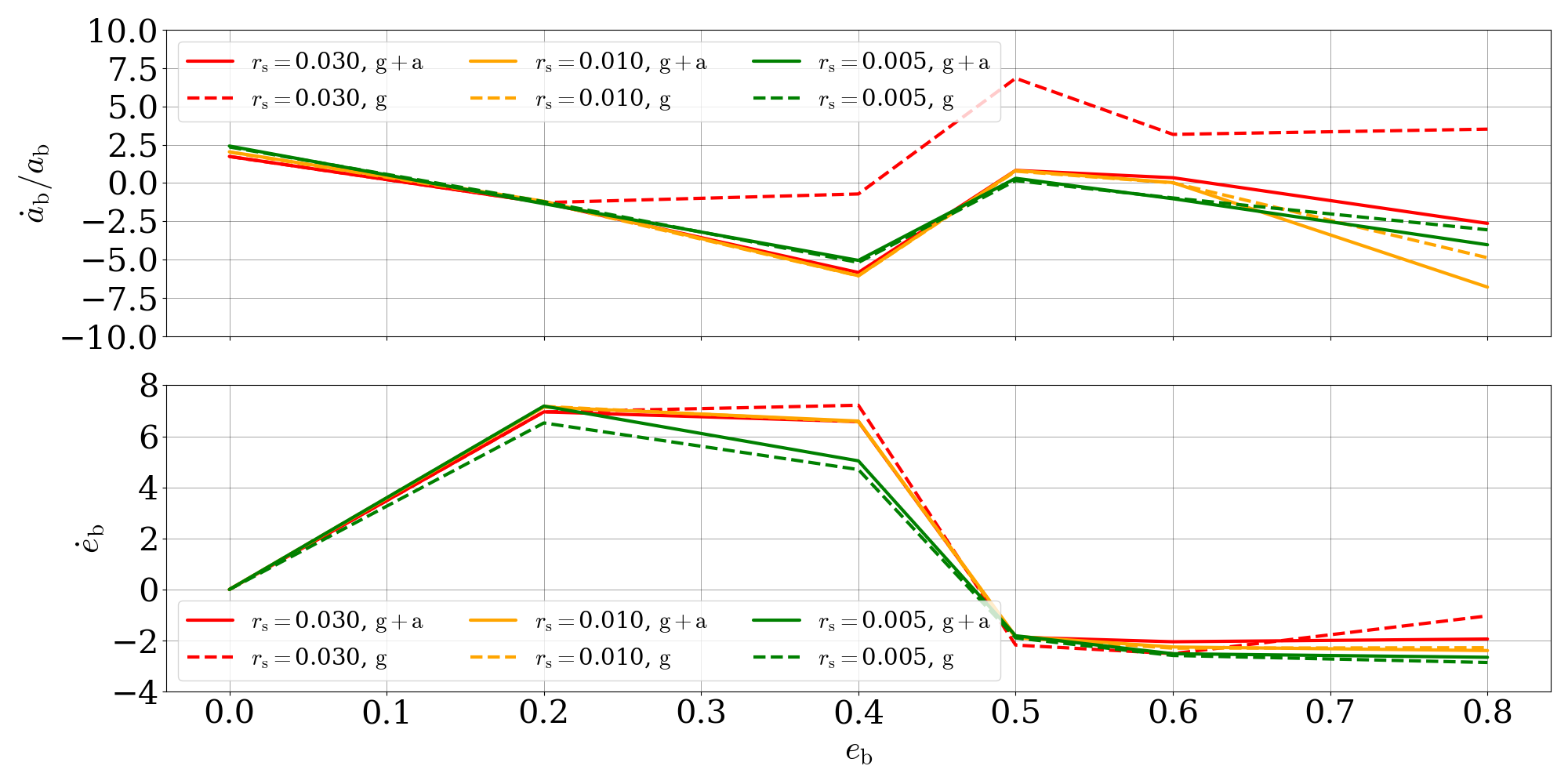}
	\caption{Semi-major axis evolution rates as a function of $e_{\rm b}$ and $r_{\rm s}$. We show the largest sink radius in red ($r_{\rm s} = 0.030 \, a_{\rm b}$), the moderate value in orange ($r_{\rm s} = 0.010\, a_{\rm b}$), and the smallest value in green ($r_{\rm s} = 0.005\, a_{\rm b}$). In addition, we show results given each sink radius value including gravity only (\texttt{g}, dashed line),  and including both gravity and accretion (\texttt{g+a}, solid line).}
	\label{fig:convergence_rs}
\end{figure*}

In Figure \ref{fig:snapshots_adot} we show a series of snapshots displaying surface density centered around the secondary ($M_{2}$) in a $e_{\rm b} = 0.8, \, q_{\rm b} = 0.1$ simulation during pericenter approach. We compare simulations of two different sink radii, from largest (top row, $r_{\rm s} = 0.03 a_{\rm b}$; our fiducial case) to smallest (second row, $r_{\rm s} = 0.005 a_{\rm b}$). In our simulation with the largest sink radius, the tidal distortion of the CSD at pericenter approach eliminates any isotropic accretion disk pattern down to the sink radius. As a result, gas streams in the cavity interact with the stripped sink particle, imparting momentum directly onto the sink. The effect on the orbital evolution is shown as a time series of $\dot{a}_{\rm b}/a_{\rm b}$ in the third row of Figure \ref{fig:snapshots_adot}: The CSD distortion coincides with large discrepancies between measures of $\dot{a}_{\rm b}/a_{\rm b}$ computed with gravity only (labeled as \texttt{g}, dashed line) and including anisotropic accretion forces  (labeled as \texttt{g+a}, solid line). These large discrepancies can lead to qualitatively different interpretations of orbital evolution (e.g. `outspiral' vs. `inspiral', see also Figure \ref{fig:convergence_rs}).

In smaller sink radius simulations, we resolve the inner region of the CSD down to $r_{\rm s} = 0.005 a_{\rm b}$. We find that the innermost region of the CSD continues to be bound to the sink, as shown in the second row of Figure \ref{fig:snapshots_adot}. Since gas remains isotropically distributed around the sink, accretion is also likely to be isotropic rather than anisotropic. As a result, we find good agreement between the \texttt{g} and \texttt{g+a} calculations (bottown row in Figure \ref{fig:snapshots_adot}), indicating negligible anisotropic accretion torques as expected. 

From the results in Figure \ref{fig:snapshots_adot} we have found that accretion torques play a non-negligible role in simulations of binaries with large enough eccentricity and sink radius. However, does this imply that accretion torques should be excluded from our fiducial simulations, to avoid numerical effects due to large sink particles? Given a small enough sink particle, towards which solution would our simulations converge?

To answer this question, we expanded our sink radius study across equal mass binaries and eccentricities in the range $e_{\rm b} = [0.0, 0.2, 0.4, 0.5, 0.6, 0.8]$. We test sink radii in the range $r_{\rm s} = [0.03  a_{\rm b}, \, 0.01  a_{\rm b}, \, 0.005 a_{\rm b}]$, and check whether the small sink radius simulations converge with the \texttt{g} or \texttt{g+a} results of our fiducial simulations, or whether the convergence points towards a different result entirely. 

In Figure \ref{fig:convergence_rs} we show a convergence study of $\dot{a}_{\rm b}$ and $\dot{e}_{\rm b}$ as a function of sink radius $r_{\rm s}$. We test sink radii in the range $r_{\rm s} = [0.03  a_{\rm b}, \, 0.01  a_{\rm b}, \, 0.005 a_{\rm b}]$, and compare calculations that include only gravitational forces (denoted as ``\texttt{g}" in the figure legend), and including both gravitational and accretion forces (denoted as ``\texttt{g+a}" in the figure legend). At the largest sink radius $r_{\rm s} = 0.030\, a_{\rm b}$, there is a significant deviation of the semi-major axis evolution when comparing the \texttt{g} and \texttt{g+a} calculation, as soon as significant binary eccentricity ($e_{\rm b} \gtrsim 0.4 $) is reached. In line with \cite{Munoz2019}, we find that anisotropic accretion torques are not negligible in highly eccentric binaries.
However, if a smaller sink radius is chosen ($r_{\rm s} \lesssim 0.010$), the \texttt{g} and \texttt{g+a} calculations yield very similar results, due to negligible contributions from anisotropic accretion torques. 

Even though anisotropic accretion torques become negligible at sufficiently small sink radii, the small sink radius simulations converge with the \texttt{g+a} result from our fiducial simulations.
 We suggest that this convergence occurs since the linear and angular momentum deposited into the sink particles by gas streams can occur in two distinct ways, depending on sink radius: 
\begin{enumerate*}
	\item when sink radii are large, the CSDs can be entirely stripped in highly eccentric binaries, since $r_{\rm s} \sim r_{\rm ER}$ (compare with equation \ref{eqn:ER}) . In this case, momentum from streams in the cavity is added to the sink particles directly through anisotropic accretion, requiring the addition of the anisotropic accretion torques in the calculation of the orbital evolution; 
	\item in simulations with small sink radii, where the sink radius is smaller than $r_{\rm ER}$ even in eccentric binaries, our simulations resolve the CSD region down to small scales. Here, momentum from streams in the cavity is added to the CSD instead of the sink particle directly, and the CSD acts as a `buffer'. The momentum transport to the CSD however changes the gas morphology near the sink particles (e.g. by shifting the disk to `lag' behind the sink), which is then reflected in the changing gravitational torques. Accretion torques are negligible, since the inner CSD is more symmetric, leading to isotropic accretion.
\end{enumerate*}

This conservation of momentum argument explains why simulations with small sink radii converge with our fiducial \texttt{g+a} simulations. The convergence suggests that our fiducial results could be applied even to binary systems such as massive black hole binaries (MBHBs), where the scale of the black holes can be comparable to our smallest sink radius $r_{\rm s} \sim 0.005 a_{\rm b}$.

We note that the binary eccentricity evolution itself is not sensitive to the numerical effects outlined above: $e_{\rm b}$ is not affected by sink treatment or inclusion of accretion torques, and as such the equilibrium eccentricities presented here are numerically robust. This is because the binary eccentricity rate of change is dominated by the CBD itself, instead of the CSD region (as discussed in section \ref{sec:eccentricity_evol}).

\section{Summary and Discussion}
\label{sec:discussion}
In this work we have investigated the orbital evolution of binaries immersed in circumbinary disks as a function of binary mass ratio and eccentricity.
We have shown that the sign and magnitude of $\dot{a}_{\rm b}$ and $\dot{e}_{\rm b}$ depends on the orbital parameter combination $[e_{\rm b}, q_{\rm b}]$ of each binary, as discussed in sections \ref{sec:semi-major_evol} and \ref{sec:eccentricity_evol}. Semi-major axis expansion is a phenomenon mostly seen in circular binaries, while eccentric binaries tend to inspiral due to gas torques. We provide numerical values of the semi-major axis evolution in Figure \ref{fig:table_abdot}, with our fiducial calculation in the left table and our calculation including only outer CBD torques on the right hand side.
The binary eccentricity also evolves due to gas torques: binaries in the mass ratio range $0.1 \leq q_{\rm b} \leq 1.0$ all evolve towards fixed equilibrium eccentricities, which are predominantly determined by gravitational interaction with gas in the CBD (as opposed to the CSDs). In Figure \ref{fig:table_ebdot}, we provide numerical values of $\dot{e}_{\rm b}$ for each binary, based on our fiducial torque calculation (\texttt{g+a}; left table), or on gravity-only calculations neglecting contributions from the cavity ($\texttt{g}_{\rm r>a}$; right table). By eye, it is easy to distinguish regions of eccentricity growth (red)  and eccentricity damping (blue). The transition between those is a diagonal across the parameter space (off-white panels), terminating at $e_{\rm b} \sim 0.5$. This diagonal represents the region of the parameter space towards which binaries evolve when interacting with CBDs. In Figure  \ref{fig:doteb_vs_eb} we show the equilibrium eccentricities found in our simulations, which grow as a function of binary mass ratio.

\subsection{Steady State: Inspiral or Outspiral?}
A key goal of our work was to predict the orbital evolution of binaries immersed in CBDs as a function of the 2-D parameter space $[e_{\rm b}, \, q_{\rm b}]$. We found that interaction with a CBD drives binaries towards equilibrium eccentricities, which increase with binary mass ratio. Binaries reach a steady state with orbital parameters $e_{\rm b, eq}, \, q_{\rm b}$ within a few $\rm{Myrs}$ (see Figure \ref{fig:eb_vs_t}), which is likely within the lifetimes of AGN disks (1-100 Myr; e.g. \citealt{HaimanHui2001, MartiniWeinberg2001, YuTremaine2002,  Khrykin2019}), or protoplanetary disks ($\gtrsim 10$ Myr; \citealt{Ronco2021}).

Once settled into their steady state eccentricities, do binaries evolve towards coalescence, or do their orbits expand? At $q_{\rm b} = 0.1$, we find that $e_{\rm b, eq} \sim 0.2$. Referring to the table in Figure \ref{fig:table_abdot}, we find that such a binary would expand due to CBD torques, and is therefore less likely to coalesce.
However, at larger mass ratios, when $q_{\rm b} \gtrsim 0.3$, all remaining equilibrium eccentricities fall into the inspiral regime (compare with the blue tiles in Figure \ref{fig:table_abdot}). Our work thus predicts that in steady state, binaries with $q_{\rm b} \lesssim 0.2$ expand, while binaries with $q_{\rm b} \gtrsim 0.3$ are driven towards coalescence by CBD torques.

\subsection{Origin of Equilibrium Eccentricities}
We have found that a non-zero equilibrium eccentricity exists for all the binaries in our simulations, and that this eccentricity increases as a function of mass ratio. This result is insensitive to the addition of accretion torques, and appears to be dominated by gas dynamics in the CBD rather than the CSDs. 

\cite{DorazioDuffell2021} suggest that the equilibrium eccentricity they find in equal mass binaries at  $e_{\rm b, eq} \sim 0.4$ is connected to the transition of CBD locking to disk precession in the same eccentricity range. In our simulation suite, we found that the equilibrium eccentricity is a function of mass ratio, and lower mass ratio binaries tend towards lower equilibrium eccentricities. However, at lower mass ratios, CBDs are locked in the entire parameter space tested (see \citealt{Siwek2023a}), with no transition to precession taking place. Our result therefore suggests that disk locking/precession regimes are unrelated to the value of the equilibrium eccentricity. However, an alternative explanation related to resonant theory stems from earlier predictions by \cite{Artymowicz1991, LubowArtymowicz1992}, and is discussed below.

Our simulations find that there is an upper limit of the equilibrium eccentricity at $e_{\rm b, eq} \sim 0.5$ for any mass ratio in the range $0.1 \lesssim q_{\rm b} \lesssim 1.0$. These results are in line with previous work by \cite{LubowArtymowicz1992}. They analyzed the eccentricity growth and damping due to resonant interactions  between binary and CBD, and found that binaries with mass ratios above $q_{\rm b} > 0.2$ evolve towards an equilibrium eccentricity $e_{\rm b,eq} \sim 0.5$. This closely mirrors the results found in this study (green line in Figure \ref{fig:doteb_vs_eb}), with gas dynamics in the cavity (included in blue and red lines in Figure \ref{fig:doteb_vs_eb})  modifying the function  $e_{\rm b,eq}(q_{\rm b})$ only slightly. 

Our work expands on recent hydrodynamic simulations by \cite{Zrake2021} and \cite{DorazioDuffell2021}, who both predicted equilibrium eccentricities for equal mass ratio binaries between $0.4 \lesssim e_{\rm b, eq} \lesssim 0.5$. The agreement between our simulations and theirs is encouraging and suggests that this result is robust to minor numerical differences in the simulations. 

\subsection{Observations of Binary Populations}
For population studies of binaries that spend part of their lifetimes evolving in tandem with CBDs, we suggest that there are `preferred' regions in the $[e_{\rm b}, q_{\rm b}]$  parameter space. These preferred regions are found along a diagonal across the $[e_{\rm b}, q_{\rm b}]$  parameter space (Figure \ref{fig:table_ebdot}), with low mass ratio systems along the lower end of eccentricities, and higher mass ratio systems near $e_{\rm b} \sim 0.5$. As mass ratios of binaries grow (slowly) due to accretion, so do their steady state eccentricities, moving them up the $[e_{\rm b}, q_{\rm b}]$  diagonal. Given a large sample of binaries, an excess of systems may be found at this diagonal due to circumbinary disk driven evolution. We discuss some examples below.

\subsubsection{MBHBs: PTAs and LISA}
The eccentricity evolution of  binaries due to interaction with accretion disks is of interest for GW and EM observations at all scales. In the case of MBHBs, orbital eccentricity may be detectable in continuous wave (CW) searches \citep{Taylor2016a} with Pulsar Timing Arrays (PTAs; \citealt{Foster1990}). At higher frequency, future space-based observatories such as the Laser Interferometer Space Antenna (LISA; \citealt{AmaroSeoane2017}) may be able to characterize the orbital eccentricities of  MBHBs and intermediate mass black hole binaries (IMBHBs) during inspiral. The residual effect of eccentricity excitation due to the CBD driven evolution of MBHBs/IMBHBs may then be measurable \citep{Zrake2021}, and would shed light on the gas-driven evolution of MBHBs and IMBHBs detected in the GW emitting regime. In future work we will apply our eccentricity calculations presented in this work to non-equal mass binaries, and constrain the orbital eccentricities at which MBHBs and IMBHBs may enter the LISA/PTA regime.

\subsubsection{LIGO Progenitors}
Circumbinary disks may form around the progenitors of compact object binaries following the late stages of common envelope evolution (CEE; \citealt{Paczynski1976}, see also \cite{Roepke2022} for a recent review). This is especially likely if the common envelope is not entirely ejected (e.g. \citealt{KashiSoker2011}), leaving a residual gas reservoir which could form a CBD. Dynamical interaction with the CBD may subsequently influence the orbital parameter evolution of the central binary, modifying properties of the growing observed population of merging compact-object binaries, such as those observed by the LIGO and Virgo gravitational-wave detectors \citep{Aasi2015, Acernese2015}.


\subsubsection{Stellar Binaries}
The Gaia Collaboration \citep{GaiaCollaboration2016} recently published a large sample of stellar binaries prompting the discovery of an excess of wide ``twin"-binaries (e.g., \citealt{ElBadry2019}). The formation of twin binaries is likely driven by circumbinary disks (e.g. \citealt{ElBadry2019,Hwang2022c}), due to the excess accretion onto the secondary. We suggest that further evidence of CBDs in stellar binary formation and evolution could be obtained using our simulations: our new results connecting binary mass ratio and eccentricity provide an opportunity to search for further evidence of circumbinary disk driven formation of stellar binaries, by statistically comparing the mass ratio and eccentricity distributions in short-period binaries.  As an example we point out the classical T Tauri star DQ Tan, which has a mass ratio close to 1 and an orbital eccentricity $e_{\rm b} =0.556$ \citep{Mathieu1997}, in agreement with our predictions. Further data analysis of the large data sets provided by Gaia can help establish the effect of circumbinary disk driven evolution on stellar binaries with more certainty.


\subsection{Caveats}
We have made several simplifications and assumptions in our simulations, which should be taken into account when interpreting our results. \begin{enumerate*}
	\item Throughout our simulations, we have modeled the viscosity using an $\alpha$-prescription and a locally isothermal equation of state, taking $\alpha = 0.1$ and the scale height of the disk $h = 0.1$. We caution that this choice of parameters may not apply to disks at all scales: AGN disks are typically assumed to be several orders of magnitude thinner \citep{Shakura1973}, while stellar disks may have lower values of $\alpha \sim 0.01$ \citep{Hartmann1998}, likely since the low ionization fraction suppresses the angular momentum transport \citep{Gammie1996} due to magnetorotational instability-driven turbulence \citep{BalbusHawley1991}.
	\item In this work we have simulated coplanar disks and binaries, without modeling any 3D effects, such as inclination angles between the disk and binary. Future work should address the question of orbital evolution (in particular eccentricity evolution) of binaries interacting with inclined disks (however, note the inclined disk simulations around circular binaries by \citealt{Moody2019}). 
\end{enumerate*}

\section{Acknowledgements}
\label{sec:ack}
MS thanks Steve Lubow, Morgan MacLeod and Ramesh Narayan for helpful discussions.
This research was supported in part by the National Science Foundation under Grant No. NSF PHY-1748958. 
This work was also supported by the Black Hole Initiative at Harvard University, which is funded by grants from the John Templeton Foundation and the Gordon and Betty Moore Foundation. 
RW is supported by the Natural Sciences and Engineering Research Council
of Canada
(NSERC), funding reference $\#$CITA 490888-16.

\section{Data Availability}
The data underlying this article will be shared on reasonable request to the corresponding author.


\twocolumn

\bibliographystyle{mnras}
\bibliography{mybib} 

\end{document}